\documentclass[12pt]{article}
                                                           
\catcode`\@=11
\renewcommand{\thefootnote}{\fnsymbol{footnote}}
\makeatletter \ifcase\@ptsize \font\teneufm=eufm10
\font\seveneufm=eufm7 \font\fiveeufm=eufm5
\font\teneusm=eusm10 \font\seveneusm=eusm7
\font\fiveeusm=eusm5 \or \font\teneufm=eufm10 scaled
\magstephalf \font\seveneufm=eufm7 \font\fiveeufm=eufm5
\font\teneusm=eusm10 scaled \magstephalf
\font\seveneusm=eusm7 \font\fiveeusm=eusm5 \or
\font\teneufm=eufm10 scaled \magstep1 \font\seveneufm=eufm7
\font\fiveeufm=eufm5 \font\teneusm=eusm10 scaled \magstep1
\font\seveneusm=eusm7 \font\fiveeusm=eusm5 \fi

\newfam\eufmfam \newfam\eusmfam \textfont\eufmfam=\teneufm
\scriptfont\eufmfam=\seveneufm
\scriptscriptfont\eufmfam=\fiveeufm
\textfont\eusmfam=\teneusm \scriptfont\eusmfam=\seveneusm
\scriptscriptfont\eusmfam=\fiveeusm

\def\frak{\ifmmode\let\next\frak@\else
 \def\next{\errmessage{Use \string\frak\space only in math
 mode}}\fi\next} \def\frak@#1{{\frak@@{#1}}}
 \def\frak@@#1{\fam\eufmfam#1} 
 \def\sh{\ifmmode\let\next\sh@\else
 \def\next{\errmessage{Use \string\sh\space only in math
 mode}}\fi\next} \def\sh@#1{{\sh@@{#1}}}
 \def\sh@@#1{\fam\eusmfam#1}

\ifcase\@ptsize \font\tenmsa=msam10 \font\sevenmsa=msam7
 \font\fivemsa=msam5 \font\tenmsb=msbm10
 \font\sevenmsb=msbm7 \font\fivemsb=msbm5 \or
 \font\tenmsa=msam10 scaled \magstephalf
 \font\sevenmsa=msam7 \font\fivemsa=msam5
 \font\tenmsb=msbm10 scaled \magstephalf
 \font\sevenmsb=msbm7 \font\fivemsb=msbm5 \or
 \font\tenmsa=msam10 scaled \magstep1 \font\sevenmsa=msam7
 \font\fivemsa=msam5 \font\tenmsb=msbm10 scaled \magstep1
 \font\sevenmsb=msbm7 \font\fivemsb=msbm5 \fi

\newfam\msafam \newfam\msbfam \textfont\msafam=\tenmsa
\scriptfont\msafam=\sevenmsa
\scriptscriptfont\msafam=\fivemsa \textfont\msbfam=\tenmsb
\scriptfont\msbfam=\sevenmsb
\scriptscriptfont\msbfam=\fivemsb

\def\Bbb{\ifmmode\let\next\Bbb@\else
 \def\next{\errmessage{Use \string\Bbb\space only in math
 mode}}\fi\next} \def\Bbb@#1{{\Bbb@@{#1}}}
 \def\Bbb@@#1{\fam\msbfam#1} \def\hexnumber@#1{\ifnum#1<10
 \number#1\else \ifnum#1=10 A\else\ifnum#1=11
 B\else\ifnum#1=12 C\else \ifnum#1=13 D\else\ifnum#1=14
 E\else\ifnum#1=15 F\fi\fi\fi\fi\fi\fi\fi}
 \def\msa@{\hexnumber@\msafam} \def\msb@{\hexnumber@\msbfam}
 \mathchardef\square="0\msa@03

\makeatother
\newcommand{\beq}{\begin{equation}}
\newcommand{\eeq}{\end{equation}}

\newcommand{\ba}{\begin{array}}
\newcommand{\ea}{\end{array}}
\newcommand{\bea}{\begin{eqnarray}}
\newcommand{\eea}{\end{eqnarray}}
\newcommand{\bean}{\begin{eqnarray*}}
\newcommand{\eean}{\end{eqnarray*}}
\newtheorem{theorem}{Theorem}[section]

\newtheorem{remark}[theorem]{Remark}

\newtheorem{proof}{Proof.}

\makeatother
 
\newcommand{\CC}{{\Bbb C}} \newcommand{\PP}{{\Bbb P}}
\newcommand{\ZZ}{{\Bbb Z}}

 \def\be{\beta}

\def\be{\begin{equation}}
\def\ee{\end{equation}}
\def\l{\label}

\def\F{{\cal F}}

\def\H{{\cal H}}

\def\K{{\cal K}}

\def\M{{\cal M}}
\def\H{{\cal H}}

\begin{document}

\begin{titlepage} 

\rightline{\tt cond-mat/9911383}
\rightline{MIT-CTP-2920}

\begin{center} 

{\Large \bf Potts model: Duality, Uniformization}

\vspace{0.333cm}

{\Large \bf and the Seiberg--Witten modulus}

\vspace{.999cm}

{\large Gaetano Bertoldi} 
\vspace{.2in}

{\it  Center for Theoretical Physics\\
        Laboratory for Nuclear Science\\ and Department of Physics\\
Massachusetts Institute of Technology\\ Cambridge, Massachusetts 02139\\ 
e-mail: bertoldi@ctp.mit.edu\\}
\vspace{.08in}

\end{center}

\vspace{.333cm}

\centerline{\large Abstract}

\vspace{0.333cm}

\noindent
The introduction of a modulus $z(K)$, 
analogous to $u=\langle tr \phi^2 \rangle$ in the 
$N=2$ SUSY $SU(2)$ gauge theory solved 
by Seiberg and Witten,
and whose defining property is the  
invariance under the symmetry and 
duality transformations of 
the effective coupling $K$,
reveals an intriguing 
correspondence between the $D=2$ Ising 
and Potts models on the square lattice.
The moduli spaces of both 
models, the spaces of 
inequivalent effective temperatures $K$,
correspond to a three--punctured sphere  
$\M_3=\PP^1(\CC) \backslash \{z=\pm1,\infty\}$.
Furthermore, in both models, the locus of Fisher 
zeroes is given by the segment joining  
$z_c=-1$ to $z_c=+1$. 

\vspace{0.333 cm}
   


\end{titlepage}
\newpage
\setcounter{footnote}{0} 
\renewcommand{\thefootnote}{\arabic{footnote}}

\section{Introduction}

Duality plays a crucial role in the latest 
developments of string theory
and has proven to be a tremendously powerful 
tool in the solution 
of non trivial $D=4$ supersymmetric field theories, 
namely the Seiberg--Witten models \cite{SW}.

In \cite{FS1} \cite{FS2}, Fendley and Saleur 
showed that various $(1+1)$--dimensional quantum impurity 
systems display an exact form of 
self--duality, analogous to the famous Kramers--Wannier duality relation 
in the $D=2$ Ising model \cite{KW}. They achieved this by expressing the 
relevant quantities in terms of contour integrals over 
certain hyperelliptic curves.

Previously, in \cite{F}, Fendley had expressed the 
magnetization in the Kondo model and the current 
in the boundary sine--Gordon model in terms of such
contour integrals, noticing the similarity with
the Seiberg--Witten results.

Fendley and Saleur also posed the question whether the duality 
relation itself is enough to solve the specific model, 
without making use of the Bethe ansatz technique. 

This problem motivates our analysis which will focus on the 
$D=2$ Ising and 
Potts models on a square lattice.
We will see that the Kramers--Wannier duality relation \cite{KW} 
and its generalization to the Potts model \cite{Potts}
naturally lead to identify a three--punctured Riemann sphere 
$\M_3=\PP^1(\CC) \backslash
\{z=\pm1,\infty\}$ as the moduli space of the given model. 

This is achieved by the introduction of a modulus
or uniformizing coordinate $z$, with the property
of being invariant under the symmetry and
duality transformations. 

\noindent
In particular 
$$
z={1+\sinh^4(2K) \over 2 \sinh^2(2K)},
$$
\be
z(K)=z(K^*)=z(-K)=z(K+i{\pi \over 2}),
\l{intro1}\ee
in the Ising case, and 
$$
z={(e^K-1)^2+q \over 2\sqrt{q}(e^K-1)},
$$
\be
z(K)=z(K^*)=z(K+2\pi i),
\l{intro2}\ee
in the Potts case,
where $K$ is the effective coupling
between neighbouring spins.

Furthermore, in terms of the appropriate modulus
$z$, the locus of Fisher zeroes  
both in the Ising and in the $q$--state Potts model
is given by the segment joining the complex temperature 
singularity $z=-1$ and the physical singularity $z=+1$.

\section{Ising model}

The Ising model is defined as follows \cite{Ising}.
At each site of a $D$--dimensional lattice, there is 
a spin variable $\sigma_i=\pm 1$ interacting with its 
nearest neighbours only through the energy
\be
\H=-J \sum_{\langle i,j\rangle} \sigma_i \sigma_j,
\l{isi1}\ee
where the symbol $\langle i,j\rangle$ denotes the sum
over nearest neighbours pairs.
The partition function $Z$ is given by 
\be
Z=\sum_{\{\sigma\}} e^{-\beta \H}=\sum_{\{\sigma\}} 
\exp\left({K\sum_{\langle i,j\rangle} \sigma_i \sigma_j}\right),
\l{isi2}\ee
where $K=\beta J$ and the sum is over all possible 
spin configurations.  

The free energy per spin $f(K)$ is given by
\be
-\beta f(K) = \lim_{N\to \infty}{\ln Z(K) \over N},
\l{isi3}\ee
Let us define
\be
F(K)\equiv-\beta f(K).
\l{isi4}\ee
In the following, we will consider a two--dimensional
square lattice with $N$ sites. 

At high temperature and correspondingly small $K$
\be
Z_{high}(K)=\sum_{\{\sigma\}} \prod_{\langle i,j\rangle} \cosh(K)
(1+\sigma_i \sigma_j \tanh(K)).
\l{isi6a}\ee
The sum over all $\sigma_i=\pm 1$ selects only 
those terms with even powers of $\sigma_i$,
while the others cancel exactly. 
We can represent each term diagrammatically
by a line connecting the sites $i$ and $j$
for each factor $\tanh(K)\sigma_i\sigma_j$.
Therefore, the non vanishing contributions 
come from all closed loops on the lattice 
\be
Z_{high}(K)=2^N (\cosh K)^{2N} \sum_{loops} 
(\tanh K)^{l}.
\l{isi6b}\ee
where $l$ denotes the length of the loop, that is
the number of bonds in the loop.
Note that the loops can be disconnected.
By (\ref{isi6b}), we see that the partition function 
is invariant under 
\be
K \to -K,
\l{isi7}\ee 
due to the fact that $l$ is even.

Likewise, at low temperature, large $K$,
a spin configuration is identified by the boundaries
of positive spin droplets in a negative spin background
or viceversa. These boundaries form loops and 
the partition function can be expressed as
\be
Z_{low}(K)=2 e^{2NK} \sum_{loops} 
e^{-2K l}.
\l{isi5}\ee
From this we easily see that under the shift
\be
K \to K+ {i \pi \over 2},
\l{tsim}\ee 
the free energy transforms
like
\be
F_{low}(K+i{\pi \over 2})=F_{low}(K)+{i \pi \over 2}.
\l{tsimf}\ee

Furthermore, by (\ref{isi6b}) and (\ref{isi5}), 
we see that if we set 
\be
e^{-2K^*}=\tanh K,
\l{KW1a}\ee 
the high--temperature
and low--temperature expansions are mapped into each other   
\be
{Z_{low}(K^*) \over 2 e^{(2NK^*)} } = 
{Z_{high}(K) \over 2^N \cosh(K)^{2N} }. 
\l{KW1b}\ee
This is the Kramers--Wannier self--duality relation \cite{KW}.
In terms of the free energy $F(K)$, (\ref{KW1b})
becomes
\be
F_{low}(K^{*}) -{1 \over 2}
\ln{\sinh(2K^*)}= F_{high}(K)- {1 \over 2}\ln{\sinh(2K)},
\l{KW1}\ee
where we used 
\be 
e^{-2K^*} =\tanh K \longleftrightarrow 
\sinh(2K^{*}) \sinh(2K) = 1.
\l{KW2}\ee

Arguing that $F(K)$ could have 
only one singularity for $K>0$, Kramers and Wannier
concluded that the transition temperature 
had to coincide with the fixed point $K^{*}(K_c)=K_c$ 
under the duality mapping 
\be
\sinh^2(2K_c) = 1 \leftrightarrow
K_c = {1 \over 2} \ln{(1+\sqrt{2})}.
\label{KW3}\ee 
Note that the duality mapping (\ref{KW2}) connects the 
low temperature region with the high 
temperature region and viceversa: in this sense it is a 
mapping between a strong coupling regime and 
a weak coupling one. Furthermore, it is involutive
\be
K^{*}(K^{*}(K)) = K.
\l{KW4}\ee

\section{The modulus $z$}

In \cite{SW}, Seiberg and Witten managed to calculate
the low--energy Wilsonian effective action 
of the $N=2$ supersymmetric $SU(2)$ 
Yang--Mills theory.
This effective action is given in terms 
of the so--called prepotential $\F(a)$,
which is a function of the vacuum 
expectation value $a$ of the scalar field 
$\phi$ contained in the supersymmetric 
multiplet. Most importantly, 
$N=2$ supersymmetry constrains
$\F(a)$ to be a polymorphic function,
that is a multivalued analytic function. 

The crucial feature is that 
the theory has a {\it moduli space} $\M$,
namely a manifold of physically 
inequivalent vacua, which is parametrized 
by the gauge--invariant 
coordinate 
\be
u=\langle \,tr \phi^2 \,\rangle.
\l{variableu}\ee
The low--energy description breaks down 
at certain points $u_c$, due to the appearance
of extra massless particles.
Furthermore, as $u$ loops around these 
singular points, $a(u)$ and its dual 
$a_D(u)={\partial \F(a) \over \partial a}$
undergo a {\it monodromy}, they transform  
into a linear 
combination of themselves
\be
\left(
\begin{array} {c}
a_D(\tilde u) \cr 
a(\tilde u)
\end{array}
\right)=
\left(
\begin{array} {cc}
A & B \cr
C & D
\end{array}
\right)
\left(
\begin{array} {c}
a_D(u) \cr 
a(u)
\end{array}
\right),
\l{2}\ee
where $\tilde u= u_c +e^{2 \pi i}(u-u_c)$ and 
$A, B, C, D$ are integers satisfying $AD-BC=1$.

These monodromies correspond to 
duality transformations relating two
theories with different effective
coupling constants  
$\tau=\partial_a^2 \F(a)$, $Im \tau >0$.. 
Eq. (\ref{2}) implies that 
\be
\tau=
{\partial a_D \over \partial a}
\rightarrow \gamma\tau={\partial \tilde a_D
\over \partial \tilde a}={A\tau +B \over C\tau+D},\quad Im \gamma\tau >0,
\ee
and 
\be
u(\gamma \tau)=u(\tau).
\l{mod3}\ee
Arguing that the theory had only 
three singularities, which could be set at
$\{ u_c= \pm 1, \infty \}$, Seiberg and Witten 
were able to determine $a(u)$ and $a_D(u)$ 
exactly in terms of period integrals of a
suitable meromorphic $1$--form 
over the two canonical homology basis cycles 
of an elliptic curve
parametrized by $u$.

These period integrals have the 
desired monodromies and are known to satisfy 
second--order linear differential equations 
called Picard--Fuchs equations. 
In particular
\be
\left( \partial_u^2 +{1 \over 4(u^2-1)}
\right) \left(
\begin{array} {c}
a_D(u) \cr 
a(u)
\end{array}
\right)=0. 
\l{lll}\ee
Note that, when $u$ loops around one of the
punctures, the differential equation does
not change because the potential is single--valued.
Therefore, $a_D(\tilde u)$ and $a(\tilde u)$
must be a linear combination of $a_D(u)$
and $a(u)$ with constant coefficients, 
as in (\ref{2}). In general, multivalued functions
with non--trivial monodromies are naturally
associated to linear differential equations 
with meromorphic potentials.
 
From Eq.(\ref{lll}), Matone 
derived a non--perturbative identity
relating $u$ and $\F(a)$ 
\be
u={\pi i}\left(\F(a)-{1 \over 2}a a_D \right),
\l{nonpertmato}\ee
which allowed to find recursion relations 
for the instanton contributions to the
prepotential \cite{Matone}.

The identity (\ref{nonpertmato}) was verified 
by instanton calculations \cite{DKM}
\cite{FT}, and it follows from
the superconformal Ward identities as well
\cite{HW}. In \cite{BMT},
Bonelli, Matone and Tonin took it as a 
starting point for a rigorous derivation 
of the Seiberg--Witten
result by reflection symmetry, 
without any assumption on the number 
of singularities \cite{BMT}.
They univocally identified the 
three--punctured sphere $\M_3=\PP^1(\CC)
\backslash \{ u_c=\pm 1, \infty \}$ as 
the moduli space of the model.

The Kramers--Wannier self--duality relation 
and its generalization in the Potts model 
imply that the physical properties 
of the system at low and high temperatures, 
equivalently high and low $K$,
are related.

Furthermore, $K$ plays the same role 
as the effective coupling constant $\tau$ in 
the Seiberg--Witten model.
Therefore, 
it is natural to look for a new variable
$z(K)$, analogous to $u(\tau)$,
that is left invariant under both the duality 
mapping (\ref{KW2}) and the symmetries (\ref{isi7}), (\ref{tsim})
\be
z(K) = z(K^{*}) = z(K + {i\pi\over 2}) = z(-K).
\l{z1}\ee
Consequently, we may regard the free energy $F(K)$ as a 
multivalued function of $z$, $\F(z)$, and
the various transformation properties 
of $F(K)$ under (\ref{isi7}), (\ref{tsim}) and (\ref{KW2}) 
will be reflected in the polymorphicity of $\F(z)$. 
In fact, performing one of these mappings is
equivalent to 
the modulus $z$ going around a 
non trivial closed path in the moduli space $\M$.
Correspondingly, the free energy $\F(z)$ will undergo 
a monodromy.

This procedure will show us that the natural setting
or moduli space
for the Ising model in two dimensions is again a 
three--punctured sphere $\M_3=\PP^1(\CC) \backslash
\{z=\pm1,\infty\}$. As we will discuss below, 
this is of course encoded in Onsager's solution 
\cite{Onsager}. We will consider 
the Potts case later.   

By (\ref{KW2}), it is natural to consider the effect of the 
various transformations on $s \equiv \sinh(2K)$. In particular, 
$s \to -s$ under both (\ref{isi7}) and (\ref{tsim}), 
and $s \to s^{-1}$ under (\ref{KW2}). We see that the maps
\be
f: s \to -s \qquad g: s \to {1 \over s},
\l{isiz}\ee
are involutive and commute
\be
f \circ f = id = g \circ g, \qquad f \circ g = g \circ f
\rightarrow (f\circ g)\circ(f \circ g)=id.
\l{isiz2}\ee
Hence the polynomial in the variable $x$
\be
P(x,s)=(x-s)(x-f(s))(x-g(s))(x-f(g(s))),
\l{isiz3}\ee 
will satisfy 
\be
P(x,s)=P(x,f(s))=P(x,g(s))=P(x,f(g(s))),
\l{isiz4}\ee 
and its coefficients will be invariant under
(\ref{isi7}), (\ref{tsim}) and (\ref{KW2}).  
We have 
\be 
P(x,s)=(x-s)(x+s)(x-{1\over s})(x+{1\over s})=
x^4 -\left({s^4+1\over s^2}\right)+1,
\l{isiz5}\ee
which implies that ${s^4+1\over s^2}$ is the quantity we are 
looking for.
We remark that there are no transformations other than 
(\ref{isi7}), (\ref{tsim}) and (\ref{KW2})
which leave $z$ unchanged.
This is a minimal choice in a sense, since no
extra spurious transformations enter the picture. For instance we can 
easily check that all the quantities:
\be
H(s,n)={1+s^{4n} \over s^{2n}},
\l{isiz6}\ee
are invariant under (\ref{isi7}), (\ref{tsim}) and (\ref{KW2}). 
But these are not the
only symmetry transformations. For example, $H(s,2)$ would be 
invariant under $s \to i s$ as well: however, we do not know 
how the free energy transforms under this map.   

Finally, we will set:
\be
z={s^4+1\over 2s^2},
\l{modulo}\ee
so that the critical temperature corresponds to $z_c(K)=1$.

We remark that there is no loss of generality in picking
$s$ instead of $\tanh(K)$ or $e^{-2K}$ as the building block. 
The invariants one finds using the latter are equivalent to $z$.

We would like to view the free energy as a 
function of $z$. To this end, we 
have to invert (\ref{modulo}) and obtain $s=s(z)$, 
which is of course a multivalued function of $z$. 
In general, by construction, for a given value of $z$
there are four different values of $s$, obtained by 
solving the equation
\be
A(s,z) = s^4 -2zs^2 + 1 = 0.
\l{Asz}\ee
However, there are certain critical values $z_c$
such that the polynomial (\ref{Asz}) has multiple roots.
Equivalently, $z_c=z(s_i)$, where $s_i$ are 
ramification points of the map $z(s)$. 
As $z$ winds around one of these critical values,
we move from one branch of $s(z)$ to another.  
This is analogous to $y^2=x$: as $x$ winds around $0$
or equivalently $\infty$, we move from $y=\sqrt{x}$ to 
$y=-\sqrt{x}$. 

If $s_0$ is a root of $A(s,z)$,
then the others will be given by $-s_0, 1/s_0, - 1/s_0$.
Therefore, we can distinguish three cases:
\begin{itemize}
\item[1.] $s_0 = -s_0 \rightarrow s_0 = 0,\infty \rightarrow z_c = \infty.$
\item[2.] $s_0 = 1/s_0 \rightarrow s_0^2 =1 \rightarrow z_c= 1.$
\item[3.] $s_0 = -1/s_0 \rightarrow s_0^2 = -1 \rightarrow z_c = -1.$
\end{itemize}
\smallskip
Hence, we have found six $s_i$, $\{\pm 1,\pm i, 0,\infty \}$, 
which correspond to three singular points, 
namely $z_c=\{\pm 1,\infty\}$. 
By (\ref{modulo}), we find
\be
s(z)=\pm \sqrt{z \pm \sqrt{z^2-1}}.
\l{isiz8}\ee
Hence, when $z$ loops around $z_c=1$ 
\be
z-1 \to e^{2\pi i}(z-1) \Rightarrow 
s(z)=\pm \sqrt{z \pm \sqrt{z^2-1}} \to
\pm \sqrt{z \mp \sqrt{z^2-1}}={1\over s(z)},
\l{isiz9}\ee  
which is equivalent
to $K \to K^*$.
On the other hand, we see that looping
around $\infty$ may correspond to either $K \to -K$
or $K \to K + i\pi/2$, since 
\be
z \to e^{2 \pi i}z \Rightarrow s(z) \sim \pm(2z)^{\pm 1}
\rightarrow \mp(2z)^{\pm 1} \sim -s(z).
\l{tdual}\ee
Indeed, as $z \to \infty$  
\be
c(z) = \cosh 2K(z) = \pm \sqrt{z+1\pm \sqrt{z^2-1}},
\l{cosh}\ee
may have the following
asymptotic behaviours
\be
c(z) \sim \pm \sqrt{2z},
\l{cosh2}\ee
or
\be
c(z) \sim \pm \sqrt{1 +{1 \over 2z}}. 
\l{cosh3}\ee
In the first case we see that
\be
z \to  e^{2 \pi i}z \Rightarrow c(z) 
\rightarrow -c(z),
\l{tdual2}\ee
which is equivalent to 
$K \to K+{i \pi \over 2}$, whereas
in the other case
\be
z \to e^{2 \pi i}z \Rightarrow c(z) \rightarrow c(z)
\l{tdual3}\ee
which is equivalent to $K \to -K$.

Let us set
\be
\F(z)=F_{high}(K(z)),
\l{effe}\ee
and
\be 
\F_D(z)=F_{low}(K(z)). 
\l{effeduale}\ee
By (\ref{KW1b}) and (\ref{KW1}), we see that 
looping around $z_c=1$
$$
\F(z) \rightarrow \F(\tilde z)=F_{high}(K(\tilde z))=
F_{high}(K^*(z))=
$$
\be
F_{low}(K(z))-\ln \sinh 2K(z) 
= \F_D(z)-\ln s(z).
\l{sdual}\ee
In a more symmetric fashion
\be
\F(z)-{1 \over 2}\ln s(z) \rightarrow
\F(\tilde z)-{1 \over 2}\ln s(\tilde z)=
\F_D(z)-{1 \over 2}\ln s(z),
\l{sdual2}\ee
where we used the fact that $s(\tilde z)=s(z)^{-1}$.
Furthermore, by (\ref{isi5}) and (\ref{tsimf}), with
$s(z)=\sqrt{z+\sqrt{z^2-1}}$
\be
z \to e^{2\pi i}z \Rightarrow \F_D(z) \rightarrow
\F_D(z)+{i \pi \over 2}. 
\l{tdual4}\ee

\section{The Uniformization Equation}\l{unifor}

We said above that this geometrical structure is 
already encoded in Onsager's solution.
Indeed, the derivative of the free energy,
the internal energy per spin $U(K)$ 
can be expressed in terms of elliptic integrals
(see for example \cite{Huang})
\be
U(K)={\partial \over \partial \beta}[\beta f(K)]=   
-J \coth2K \left[1+k_1'
{2 \over \pi} \K(k_1) \right],
\l{ident1}\ee
\medskip
where $\K(k_1)$ is the complete elliptic integral 
of the first kind 
\be
\K(k_1)=\int^{\pi \over 2}_0 
{ d\phi \over \sqrt{1-k_1^2 \sin^2 \phi}}
\l{k1}\ee
and
\be
k_1={2 s \over s^2+1} \qquad k_1'={s^2-1\over s^2+1}\qquad 
k_1^2+(k_1')^2=1.
\l{k3}\ee
\medskip
In terms of $F(K(z))=\F(z)$, using (\ref{isi4})(\ref{ident1}) and
$K=\beta J$
\be
U(K)=
-{\partial K \over \partial \beta}\partial_K F(K)=
-J{\partial z \over \partial K}\partial_z \F(z),
\ee
which implies, after some algebra, that
\be
\partial_z \F(z)={\sigma \over 4\sqrt{z^2-1}}+ 
{1 \over \pi (z+1)}\K(k_1),
\l{identif}\ee
where $\sigma=\pm 1 \leftrightarrow 
s^2=z+\sigma\sqrt{z^2-1}$.

Furthermore, note that $\sqrt{z-1}\,\K(k_1(z))$ is a  
solution of the {\it uniformization
equation} for the three--punctured sphere
$\M_3=\PP^1(\CC) \backslash \{z=\pm1,\infty\}$
\be
\left[\partial^2_z + {3+z^2 \over 4(z^2-1)^2}\right]
\left(
\begin{array} {c}
\psi_D(z) \cr 
\psi(z)
\end{array}
\right)
=0.
\l{unif1}\ee
Therefore, by (\ref{identif}), $\F(z)$
solves the following third-order equation
\be
\partial^3_z \F(z) + {3z-1 \over z^2-1}\partial_z^2 \F(z) 
+{2z+1 \over 2(z+1)^2 (z-1) }\partial_z \F(z)
-{\sigma \over 8} { z+1 \over (z^2-1)^{5 \over 2}}=0.
\l{PicFuchs1}\ee

Let us denote by $H$ the Poincar\'e upper half plane
$$
H=\{\tau \in \CC \mid 
Im\,\tau > 0 \}.
$$
The uniformization theorem tells us that the
universal covering of the 
three--punctured sphere $\M_3$ is $H$.
Indeed, there exists a holomorphic covering map
$$
J_H: H \rightarrow \M_3
$$
\be
\tau \mapsto J_H(\tau)=z(\tau),
\l{lala1}\ee
such that
\be
z(\gamma \tau)=z(\tau) \iff \gamma \in \Gamma(2),
\l{lala2}\ee
where $\Gamma(2)$
is a subgroup of $PSL(2,\ZZ)$ defined by
$$
\Gamma(2)=\left\{ 
\left(
\begin{array} {cc}
a & b \cr
c & d
\end{array}
\right), ad-bc=1,\, a,d \equiv 1 \, mod\, 2, 
b,c \equiv 0 \, mod\, 2 \right\}\,,
$$
and
\be
\gamma \tau={a\tau + b \over c\tau+d}
\ee
Hence, $\M_3 \cong H/\Gamma(2)$.

The ratio of two linearly independent solutions of 
(\ref{unif1}) gives the inverse of the uniformization 
map up to an overall $PSL(2,\CC)$ transformation 
$$
J_H^{-1}: \M_3 \to H
$$ 
\be
\tau(z) =J_H^{-1}(z)={\psi_D(z) \over \psi(z)}.
\l{unif2}\ee
The function $\tau(z)$ is polymorphic: when 
$z$ loops around one of the punctures,
$\{\pm1,\infty\}$, $\psi_D(z)$ and $\psi(z)$ undergo 
a monodromy
\be
\left(
\begin{array} {c}
\psi_D(\tilde z) \cr 
\psi(\tilde z)
\end{array}
\right)=
\left(
\begin{array} {cc}
a & b \cr
c & d
\end{array}
\right)
\left(
\begin{array} {c}
\psi_D(z) \cr 
\psi(z)
\end{array}
\right),
\l{monodr1}\ee
and correspondingly
\be
\tau(\tilde z)= {a \tau(z)+b \over c\tau(z)+d}, 
\qquad
\left(
\begin{array} {cc}
a & b \cr
c & d
\end{array}
\right) \in \Gamma(2),
\l{monodr2}\ee
in accordance with (\ref{lala2}).

\section{Potts model}

The Potts model \cite{Potts} generalizes the Ising model 
in the sense that each spin $s_i$ can have $q$ values,
$(1,2,\ldots,q)$, and the Hamiltonian reads
\be
\H =-J \sum_{\langle i,j\rangle} \delta_{s_i s_j}.
\l{pot1}\ee
The partition function is given by
\be 
Z(K)=\sum_{\{\sigma\}} e^{-\beta \H}=\sum_{\{\sigma\}} 
\exp\left( {K\sum_{\langle i,j\rangle} \delta_{s_i s_j}}\right),
\l{pot2}\ee
where as usual $K=\beta J$.
Although the problem of finding the exact free energy 
for the square lattice $D=2$ Potts model has not been
solved, some exact results are known. In particular,
there exists the analogue of the Kramers--Wannier 
duality relation \cite{Potts}, which determines
the critical temperature
\be
K_c=\ln(1+\sqrt{q}).
\l{crit1}\ee
For $q=3,4$ the phase transition is second order,
while for $q>4$ the transition is first order and the latent 
heat is known exactly \cite{Baxter1}.
Furthermore, for $q>4$ the spontaneous magnetization \cite{Baxter2}
and the correlation length at $T_c$ \cite{BuffWall} are also known. 
Finally, for $q=3,4$ the critical exponents are given exactly 
\cite{denNijsetal}. For a comprehensive review see \cite{Wu}.  

Let us define
\be
F(K)\equiv\lim_{N\to \infty} {\ln Z(K) \over N}.
\l{pot3}\ee
The high temperature expansion reads
\be
Z_{high}(K)=q^N C(K)^{2N} \sum_{loops} 
T(K)^l,
\l{potts1}\ee
where 
\be
C(K)={e^K+q-1 \over q},\qquad T(K)={e^K-1 \over e^K+q-1},
\l{potts2}\ee
and $l$ denotes the length of the loop.
Conversely, for the low temperature expansion we have
\be
Z_{low}(K)=q e^{2KN} \sum_{loops} e^{-Kl}.
\l{potts3}\ee
Note that, in contrast to the Ising model,
the length $l$ can be odd. For instance,
consider $q=3$, say blue green and red spins. 
In the low temperature expansion there are 
excitations corresponding to one green and 
one red spin next to each other in a background
of blue spins. In this case $l$ is seven. 
Therefore, the analogue of (\ref{tsim}) is
\be
K \to K + 2\pi i.
\l{potts4}\ee 
Under this shift, by (\ref{potts3})
\be
F_{low}(K+2 \pi i)=F_{low}(K)+ 4 \pi i.
\l{potts3b}\ee
The generalization of the Kramers--Wannier relation 
for the Potts model is \cite{Potts}
\be
{Z_{high}(K^*) \over q^N C(K^*)^{2N}}={Z_{low}(K) \over q e^{2KN}},
\l{pottskw1}\ee
implying that
\be
F_{high}(K^*)-\ln q -2 \ln C(K^*) = F_{low}(K) - 2K,
\l{pottskw2}\ee
where
\be
e^{-K^*}=T(K)={e^K-1 \over e^K+q-1} \leftrightarrow
(e^{K^*}-1)(e^K-1)=q.
\l{pottskw3}\ee
Therefore, we see that the critical temperature 
is indeed given by (\ref{crit1}).
By virtue of (\ref{pottskw3}), we can rewrite
Eq.(\ref{pottskw2}) in a more symmetrical way,
namely
\be
F_{high}(K^*)-\ln(e^{K^*}-1)=F_{low}(K)-\ln(e^K-1).
\l{pottskw4}\ee
As before, the mapping $K^*(K)$ is involutive.

The simplicity of transformation (\ref{potts4})
makes it easier to find a variable invariant under 
this mapping and (\ref{pottskw3}).
We can basically choose either 
\be
w(K)=T(K)T(K^*)=e^{-K} e^{-K^*}={e^K-1 \over e^K(e^K+q-1)}, 
\l{potts5}\ee
or
\be
v(K)=T(K)+T(K^*),
\l{potts6}\ee
which are related by (\ref{pottskw3}), in particular
\be
v(K)=(1-q)w(K)+1.
\l{potts7}\ee
Solving Eq.(\ref{potts5}) we find
\be
e^K=
{(1-q)w+1 \pm (1-q) \sqrt{(w-{1 \over (1 + \sqrt{q})^2})
(w-{1 \over (1 - \sqrt{q})^2}) } \over 2w}.
\l{wek}\ee
By construction, there are two values of $w$ where the 
solutions of (\ref{wek}) coincide. 
These critical values $w_c=w(K_c)$
are given by
\be
w_1={1 \over (1 +\sqrt{q})^2},\quad w_2={1 \over (1-\sqrt{q})^2},
\l{potts9}\ee
where $w_1$ corresponds to the physical singularity
and $w_2$ to the complex temperature singularity.
These are the 
fixed points of the duality transformation (\ref{pottskw3}).
However, note that $F(K)$ is singular as 
$K \to 0,\infty$
\be
K \to \infty \Rightarrow F_{low}(K) \sim 2K. 
\l{potts8}\ee
Hence, we argue that there is a further singularity at
$w(K=0)=w(K=\infty)=0$.

In order to show the close relationship
with the Ising model, we will perform a M\"obius 
transformation on $w$ so that the critical points 
coincide with $\{\pm 1,\infty\}$. 

The linear fractional transformation that maps $w_1$ to $1$,
$w_2$ to $-1$ and $w=0$ to $\infty$ is given by
\be
z={1-(1+q)\,w \over 2 \sqrt{q}\, w } \iff
w = {1 \over 2 \sqrt{q}\, z +(1+q)}.
\l{zpotts}\ee
Hence
\be
z={(e^K-1)^2+q \over 2\sqrt{q}(e^K-1)},
\l{potts10}\ee
and
\be
e^K= 1+\sqrt{q}\,( z \pm \sqrt{z^2-1}).
\l{zek}\ee 
By (\ref{zek}), we see that looping around 
$z_c=1$, $K$ goes to $K^*$
\be
K(\tilde z)=\ln({1+\sqrt{q}\,( z \mp \sqrt{z^2-1})})
\rightarrow (e^{K(\tilde z)}- 1)(e^{K(z)}- 1)=q \rightarrow
K(\tilde z)=K^*(z),
\l{potts10a}\ee
where $\tilde{z}-1=e^{2\pi i}(z-1)$.
Furthermore, choosing the plus sign in (\ref{zek}),
as $z$ loops around 
$\infty$
\be
z \to e^{2 \pi i}z \Rightarrow K \sim \ln z \rightarrow
\ln z + 2 \pi i = K + 2 \pi i,
\l{potts11}\ee 
that is we retrieve the symmetry (\ref{potts4}).

\section{Fisher zeroes}

In \cite{Fisher}, Fisher emphasized the role 
of the zeroes of the partition function in the complex 
temperature plane in the study of phase transitions.
In particular, using Kaufman's expression of $Z(K)$ for 
the two--dimensional Ising model \cite{Kauf}, he showed
that, in the thermodynamic limit, 
the zeroes are dense on two circles in the $\tanh K$ or 
equivalently $e^{-K}$ plane given by
\be
\tanh K = 1 +\sqrt{2}e^{i \theta}, \quad
\tanh K = -1 +\sqrt{2}e^{i \theta.}
\l{fish1}\ee
These are the antiferromagnetic and ferromagnetic circles 
respectively and they are related by the map $K \to -K$. 

In terms of $z$, both circles correspond to the segment
joining $z_c=-1$ to $z_c=+1$.
By (\ref{fish1}), we have
$$
s^2(\theta)=\left({2 \tanh K(\theta)\over 1-\tanh^2 K(\theta)}
\right)^2=\left({\pm 1 + \sqrt{2} e^{i \theta} \over e^{i2 \theta}
\pm \sqrt{2}e^{i \theta} }\right)^2=
$$
$$
\left({\sqrt{2}\pm e^{-i \theta} \over
\sqrt{2}\pm e^{i \theta} }\right)^2.
$$
Then
$$
\bar{s}^2(\theta)={1 \over s^2(\theta)},
$$
which implies that
$$
z(\theta)={1 \over 2}\left(s^2(\theta)+{1 \over s^2(\theta)}
\right)=
\bar{z}(\theta).
$$
Therefore, the two circles are constrained on the real axis
and 
$$
z(\theta)={1\over 2}\left[\left({\sqrt{2}\pm \cos\theta\mp 
i \sin\theta \over \sqrt{2}\pm \cos\theta\pm i \sin\theta}
\right)^2 + c.c. \right]=
{1\over 2}\left[{(\sqrt{2}\pm \cos\theta\mp 
i \sin\theta)^4 \over ((\sqrt{2}\pm \cos\theta)^2 +\sin^2\theta)^2}
+ c.c. \right]=
$$
$$
{(\sqrt{2}\pm \cos\theta)^4 
-6(\sqrt{2}\pm \cos\theta)^2\sin^2\theta+\sin^4\theta \over
((\sqrt{2}\pm \cos\theta)^2 +\sin^2\theta)^2},
$$
implying that $|z(\theta)|\leq 1$, 
as it can be easily checked.
Hence, since both $z_c=-1$ and $z_c=1$ are in the image,
$z(\theta)$ maps the interval $[0,2\pi]$
into the interval $[-1,+1]$.

In \cite{Martin}\cite{MRamm}, Martin, Maillard and Rammal 
conjectured that the Fisher zeroes in the 
two--dimensional $q$--state Potts model lie on a circle
in the complex $e^{-K}$ plane given by
\be
e^{-K(\theta)}=-{1\over q-1}+{\sqrt{q}\over q-1}e^{i\theta}.
\l{circle1}\ee
Actually, they conjectured that the locus was given by the condition
\be
e^{-K^*}=e^{-\bar{K}},
\l{fish4}\ee 
where $K^*$ is the dual of $K$ and $\bar{K}$ denotes
the complex conjugate of $K$. This 
yields precisely (\ref{circle1}).
Note also that setting $q=2$ recovers the ferromagnetic circle
of the Ising model. Numerical investigations
for small lattices at $q=3$ and $q=4$ \cite{Martin}\cite{MRamm} 
provided evidence for (\ref{circle1}).
Further progress was made in \cite{MatShrock},
for $3\leq q \leq 8$.
In \cite{CHW}, on the basis of numerical results
on small lattices for $q \leq 10$, it was conjectured 
that for finite lattices with self--dual boundary conditions,
and for other boundary conditions in the thermodynamic limit,
the zeroes in the ferromagnetic regime are on the above circle.
This conjecture was actually proved for infinite $q$
in \cite{WRHMHC}. 
 
In \cite{Kenna}, using a general result concerning 
the partition function zeroes of models 
displaying first order phase transition 
obtained by Lee \cite{Lee}, Kenna proved 
that the locus for $q>4$ is indeed given by 
(\ref{circle1}). His argument relies on the fact
that the thermodynamic limit and the application 
of the duality transformation (\ref{pottskw3}) 
to the Fisher zeroes should commute.

We will now show that, in terms of the appropriate 
modulus $z$, the locus of the Fisher zeroes
for the $q$--state Potts model (\ref{circle1}) still 
corresponds to the 
segment joining $z_c=-1$ to $z_c=+1$.
   
First, note that the locus is equivalent 
to 
\be
e^{K(\theta)}=1+\sqrt{q}e^{i\theta}.
\l{circle2}\ee
Indeed, $(e^K)^{-1} \to e^K$ is a
M\"obius transformation and thus it maps 
circles into circles. Furthermore,
Eq.(\ref{circle1}) implies that
$$
e^{K(\theta)}={q-1 \over \sqrt{q}e^{i \theta}-1}\Rightarrow
e^{K(\theta)}-1={\sqrt{q}(\sqrt{q}-e^{i \theta})\over
e^{i \theta}(\sqrt{q}-e^{-i \theta})}\Rightarrow
$$
\be
|e^{K(\theta)}-1|^2=q.
\l{circle2b}\ee
Finally, by (\ref{zek}), we immediately recognize
that (\ref{circle2})  
corresponds to $z(\theta)=\cos\theta$
\medskip
\be
z \pm \sqrt{z^2-1} =e^{i\theta} \Rightarrow
z^2-2z e^{i \theta} +e^{2i\theta}=z^2-1 \Rightarrow
z(\theta)=\cos\theta.
\l{circle3}\ee
Therefore, as in the case of the Ising model,
the locus of Fisher zeroes for the Potts model
corresponds to the 
segment joining the complex temperature singularity
$z_c=-1$ to the physical singularity $z_c=+1$.

\section{Conclusions}

In summary, we have seen that 
the introduction of a proper modulus
or uniformizing coordinate $z$, which is given by
\medskip
\be
z={1+\sinh^4(2K) \over 2 \sinh^2(2K)},
\l{finale1}\ee
in the Ising case, and 
\medskip
\be
z={(e^K-1)^2+q \over 2\sqrt{q}(e^K-1)},
\l{finale2}\ee  
in the Potts case,
unveils an intriguing correspondence
between the $D=2$ Ising and Potts models. 
Indeed, their moduli spaces are 
both equivalent to the
three--punctured sphere 
\be
\M_3=\PP^1(\CC) \backslash \{z=\pm1,\infty\}.
\l{finale3}\ee
The moduli (\ref{finale1}) and (\ref{finale2})
are characterized by the invariance
under the symmetry and duality
transformations (\ref{isi7})(\ref{tsim})(\ref{KW1a}) 
and (\ref{potts4})(\ref{pottskw3})
respectively. 

This connection 
is further strengthened by the fact that
the locus of Fisher zeroes in both models
is given by the segment joining
$z_c=-1$ to $z_c=+1$.

\medskip
\medskip
\medskip
\noindent
{\bf Acknowledgements}. It is a pleasure to thank B. Feng,
L. Griguolo, M. Matone, G. Migliorini and W. Taylor for 
stimulating discussions and M. Matone and W. Taylor
for a careful reading
of the manuscript. 
I would like to thank  
A.J. Guttmann and I. Jensen for their 
valuable comments on a previous version
of this paper and for sharing a 
draft of their work \cite{GuttJens}. 

This work is supported in part by D.O.E.
cooperative agreement DE--FC02-94ER40818 and by INFN ``Bruno Rossi''
Fellowship. I also thank S.I.F. (``Societ\'a  Italiana di Fisica'')
for the ``Alberto Frigerio'' Prize.

\end{document}